\documentclass{emulateapj}

\usepackage{color}
\usepackage{multirow}
\usepackage{longtable}

\usepackage{natbib}

\shorttitle{Arizona CDFS Environment Survey (ACES)}
\shortauthors{Cooper et al.}

\begin{document}


\title{The Arizona CDFS Environment Survey (ACES): A Magellan/IMACS 
Spectroscopic Survey of the Chandra Deep Field South\footnotemark[*]}


\author{
Michael C.\ Cooper\altaffilmark{1,\dag,2,\ddag},
Renbin Yan\altaffilmark{3},
Mark Dickinson\altaffilmark{4},
St\'{e}phanie Juneau\altaffilmark{2,5}, 
Jennifer M.\ Lotz\altaffilmark{6}, 
Jeffrey A.\ Newman\altaffilmark{7}, 
Casey Papovich\altaffilmark{8},  
Samir Salim\altaffilmark{9}, 
Gregory Walth\altaffilmark{2}, 
Benjamin J.\ Weiner\altaffilmark{2}, 
Christopher N. A. Willmer\altaffilmark{2}
}

\footnotetext[*]{This paper includes data gathered with the 6.5 meter
  Magellan Telescopes located at Las Campanas Observatory, Chile.}

\altaffiltext{1}{Center for Galaxy Evolution, Department of Physics
  and Astronomy, University of California, Irvine, 4129 Frederick
  Reines Hall, Irvine, CA 92697, USA; m.cooper@uci.edu}

\altaffiltext{\dag}{Hubble Fellow}

\altaffiltext{2}{Steward Observatory, University of Arizona, 933 N.\
  Cherry Avenue, Tucson, AZ 85721, USA}

\altaffiltext{\ddag}{Spitzer Fellow}

\altaffiltext{3}{Center for Cosmology and Particle Physics,
  Department of Physics, New York University, 4 Washington Place, New
  York, NY 10003, USA}

\altaffiltext{4}{National Optical Astronomy Observatory, 950 North
  Cherry Avenue, Tucson, AZ 85719}

\altaffiltext{5}{Laboratoire AIM, CEA/DSM-CNRS-Universit\'{e}́ Paris
  Diderot, Irfu/Service dAstrophysique, CEA-Saclay, Orme des
  Merisiers, 91191 Gif-sur-Yvette Cedex, France}

\altaffiltext{6}{Space Telescope Science Institute, 3700 San Martin
  Drive, Baltimore, MD 21218, USA}

\altaffiltext{7}{Pittsburgh Particle Physics, Astrophysics, and
  Cosmology Center (PITT-PACC), Department of Physics and Astronomy,
  University of Pittsburgh, 3941 O'Hara Street, Pittsburgh, PA 15260,
  USA}

\altaffiltext{8}{Department of Physics and Astronomy, Texas A\&M
  University, College Station, TX 77845, USA}

\altaffiltext{9}{Department of Astronomy, Indiana University,
  Bloomington, IN 47404, USA}

\begin{abstract}

  We present the Arizona CDFS Environment Survey (ACES), a
  recently-completed spectroscopic redshift survey of the Chandra Deep
  Field South (CDFS) conducted using IMACS on the Magellan-Baade
  telescope. In total, the survey targeted $7277$ unique sources down
  to a limiting magnitude of $R_{\rm AB} = 24.1$, yielding $5080$
  secure redshifts across the $\sim \! 30^{\prime} \times 30^{\prime}$
  extended CDFS region. The ACES dataset delivers a significant
  increase to both the spatial coverage and the sampling density of
  the spectroscopic observations in the field. Combined with
  previously-published, spectroscopic redshifts, ACES now creates a
  highly-complete survey of the galaxy population at $R < 23$,
  enabling the local galaxy density (or environment) on relatively
  small scales ($\sim \! 1$ Mpc) to be measured at $z < 1$ in one of
  the most heavily-studied and data-rich fields in the sky. Here, we
  describe the motivation, design, and implementation of the survey
  and present a preliminary redshift and environment catalog. In
  addition, we utilize the ACES spectroscopic redshift catalog to
  assess the quality of photometric redshifts from both the COMBO-17
  and MUSYC imaging surveys of the CDFS.

\end{abstract}


\section{Introduction}
\label{sec_intro}

\setcounter{footnote}{3}

Building upon the initial X-ray observations of \citet{giacconi01,
  giacconi02}, the Chandra Deep Field South (CDFS, $\alpha = 03^{\rm
  h}32^{\rm m}25^{\rm s}$, $\delta = -27^{\circ}49^{\rm m}58^{\rm s}$)
has quickly become one of the most well-studied extragalactic fields
in the sky with existing observations among the deepest at a broad
range of wavelengths (e.g., \citealt{giavalisco04}; \citealt{rix04};
\citealt{lehmer05}; \citealt{quadri07}; \citealt{miller08};
\citealt{padovani09}; \citealt{cardamone10}; \citealt{xue11};
\citealt{damen11}; Elbaz et al.\ in prep). In the coming years, this
status as one of the very deepest multiwavelength survey fields will
be further cemented by the ongoing and upcoming extremely-deep
observations with {\it Spitzer}/IRAC and {\it HST}/WFC3-IR as part of
the Spitzer Extended Deep Survey (SEDS, PI G.\ Fazio) and the Cosmic
Assembly Near-IR Deep Extragalactic Legacy Survey (CANDELS,
\citealt{grogin11, koekemoer11}), respectively.

Despite the large commitment of telescope time from both space- and
ground-based facilities devoted to imaging the CDFS, spectroscopic
observations in the field have generally lagged those in other,
comparably-deep extragalactic survey fields. For example, in the
Extended Groth Strip, another deep field targeted by the SEDS and
CANDELS programs, the DEEP2 and DEEP3 Galaxy Redshift Surveys
(\citealt{davis03, davis07, cooper11, cooper12, newman12}; Cooper et
al.\ in prep; see also \citealt{weiner06}) have created a vast
spectroscopic database, achieving an impressive $\sim \!  60\%$
redshift completeness down to $R_{\rm AB} = 24.1$ across more than
$0.2$ square degrees and $\gtrsim \! 40\%$ completeness covering a
broader area of $\sim \! 0.5$ square degrees down to the same
magnitude limit. Similarly, there have been a variety of spectroscopic
efforts in the GOODS-N field including the Team Keck Redshift Survey
\citep[TKRS,][see also \citealt{cooper11}]{wirth04} in addition to the
independent work of various groups \citep[e.g.,][]{lowenthal97,
  phillips97, cohen00, dawson01, cowie04, treu05, reddy06,
  barger08}. Together, these datasets provide spectroscopic redshifts
for $> \! 90\%$ of sources down to $z_{\rm F850LP} = 23.3$
\citep{barger08}.

These large spectroscopic surveys add significant scientific utility
to the associated imaging datasets, making the photometric constraints
much more powerful.  For example, spectroscopic redshifts allow
critical rest-frame quantities to be derived with increased
precision. Furthermore, only through spectroscopy can assorted
spectral and dynamical properties (such as the strengths and velocity
widths of emission and absorption lines) be measured --- in this
regard, the spectral databases provided by surveys such as DEEP2,
DEEP3, and TKRS are especially powerful due to their uniform spectral
range and resolution. Finally, only with the combination of high (and
relatively uniform) sampling density, spatial coverage across a
modestly-sized field (e.g., $\gtrsim \! 0.05$ square degrees), and
moderately high-precision spectroscopic redshifts (i.e., $\sigma_{z} <
0.01$) can the local galaxy density (or ``environment'') be measured
across a broad and continuous range of environments \citep{cooper05,
  cooper07}.

In contrast to the EGS and GOODS-N fields, the spectroscopic redshift
completeness across the extended $30^{\prime} \times 30^{\prime}$ area
of the CDFS is modest, $\lesssim \! 25\%$ down to a limiting magnitude
of $R_{\rm AB} = 23$ and $\lesssim \! 20\%$ at $R_{\rm AB} < 24$,
despite some significant spectroscopic efforts in the field
\citep[e.g.,][]{lefevre04, szokoly04, vanzella05, vanzella06,
  vanzella08, mignoli05, ravikumar07, popesso09, balestra10,
  silverman10}.\footnote{The PRIMUS program \citep{coil11} has
  collected spectra for a substantial number of sources in a larger
  area surrounding the CDFS. However, the relatively low resolution
  ($R = \lambda/\Delta \lambda \sim 30$) of the prism-based
  spectroscopy limits the utility of the PRIMUS dataset for detailed
  studies of spectral properties (e.g., emission-line equivalent
  widths) and small-scale environment (e.g., on group scales).}
Notably, many of these existing spectroscopic programs have focused
their efforts on the smaller GOODS-S region and/or targeted
optically-faint, higher-redshift $(z > 1.5)$ sources
\citep[e.g.,][]{dickinson04, doherty05, roche06, vanzella09}.

With the goal of creating a highly-complete redshift survey at $z < 1$
in the CDFS, the Arizona CDFS Environment Survey (ACES) utilized the
Inamori-Magellan Areal Camera and Spectrograph
\citep[IMACS,][]{dressler11} on the Magellan-Baade telescope to
collect spectra of more than $7000$ unique sources across a $\sim \!
30^{\prime} \times 30^{\prime}$ region centered on the CDFS. In
Sections \ref{sec_design} and \ref{sec_redux}, we describe the design,
execution, and reduction of the ACES observations, with a preliminary
redshift and environment catalog presented in Sections \ref{sec_data}
and \ref{sec_environ}. Finally, in Section \ref{sec_future}, we
conclude with a discussion of future work related to ACES. Throughout,
we employ a Lambda cold dark matter ($\Lambda$CDM) cosmology with $w =
-1$, $\Omega_{\rm m}$ = 0.3, $\Omega_{\Lambda} = 0.7$, and a Hubble
parameter of $H_{0} = 100\ h$ km s$^{-1}$ Mpc$^{-1}$. All magnitudes
are on the AB system \citep{oke83}, unless otherwise noted.

\begin{figure}
\centering
\plotone{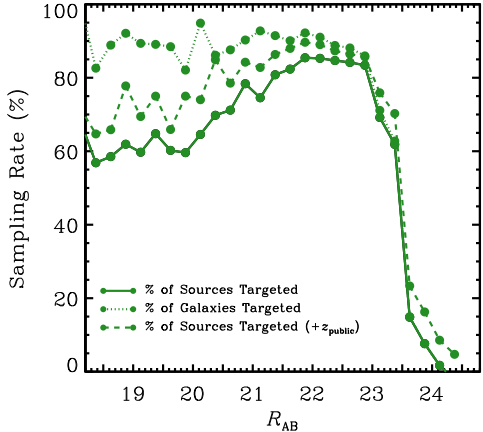}
\caption{The ACES target sampling rate as a function of $R$-band
  magnitude across the entire $\sim \! 30^{\prime} \times 30^{\prime}$
  COMBO-17/CDFS footprint. The target sampling rate is defined as the
  percentage of objects at a given $R$-band magnitude in the COMBO-17
  photometric catalog that were observed by ACES. The dotted and
  dashed lines show the sampling rate when only considering sources
  classified as non-stellar in the COMBO-17 catalog$^{5}$ (dotted) and
  when accounting for sources with a published spectroscopic redshift
  in the literature (dashed -- see \S \ref{sec_design}). At bright
  magnitudes, $R_{\rm AB} < 23$, ACES brings the target sampling rate
  in the CDFS to $> \! 80\%$.}
\label{fig_sampling_rate}
\end{figure}

\stepcounter{footnote}\footnotetext{To select the galaxy population,
  all sources classified as ``Star'' or ``WDwarf'' in the COMBO-17
  photometric catalog of \citet{wolf04} are excluded.}

\vspace*{0.2in}

\section{Target Selection and Slitmask Design}
\label{sec_design}

The ACES target sample is drawn from the COMBO-17 photometric catalog
of \citet[][see also \citealt{wolf01a,wolf08}]{wolf04}. The primary
spectroscopic sample is magnitude limited at $R_{\rm AB} < 23$, plus a
significant population of fainter sources down to $R_{\rm AB} =
24.1$. Altogether, ACES observations spanned four separate observing
seasons (2007B -- 2010B), with the details of the target-selection
algorithm and slitmask-design parameters varying from year to
year. Here, we highlight the critical elements of the composite target
population and slitmasks, including any significant variations from
mask to mask.

As stated above, the primary target sample for ACES was selected
according to an $R$-band magnitude limit of $R_{\rm AB} <
23$. Selecting in $R$ (versus a redder passband such as $I$ or $K$)
ensures the highest possible signal-to-noise ratio in the continuum of
the resulting optical spectra, and thus a high redshift-success rate
for the survey. The given magnitude limit was adopted to enable a high
level of completeness to be reached across the entire CDFS
area. Moreover, at $z \sim 1$, the $R = 23$ limit reaches $L^{*}$
along the red sequence and $1$ magnitude fainter than $L^{*}$ in the
blue cloud population \citep{willmer06}, thereby enabling ACES to
probe the systems that dominate the galaxy luminosity density and
global star-formation rate at $z < 1$. As shown in Figure
\ref{fig_sampling_rate}, ACES is highly-complete at $R < 23$,
achieving a targeting rate of $\gtrsim \! 80\%$ across the extended
CDFS.

In addition to the main $R < 23$ target sample, we prioritized
$70\mu$m sources detected as part of the Far-Infrared Deep
Extragalactic Legacy (FIDEL) Survey (PI: M.~Dickinson), which
surveyed the CDFS to extremely deep flux limits at both $24\mu$m and
$70\mu$m with {\it Spitzer}/MIPS \citep{magnelli09, magnelli11}. In
selecting the optical counterparts to the $70\mu$m sources, we
utilized a fainter magnitude limit of $R = 24.1$, targeting multiple
optical sources in cases for which the identity of the optical
counterpart was ambiguous. In total, ACES prioritized $529$ sources as
$70\mu$m counterparts, with $\sim \! 80\%$ of these sources also
meeting the $R < 23$ primary magnitude limit for the survey. For the
$70\mu$m-selected target population, our redshift success rate is
quite high ($\sim \!  80\%$ versus $\sim \! 70\%$ for the full target
population).

As a filler population in the target-selection process, we also
included (with a lower selection probability) all sources down to the
secondary magnitude limit of $R < 24.1$. These fainter,
optically-selected sources comprise roughly $30\%$ of the total unique
target sample (i.e, $\sim \! 2500$ sources). The $R = 24.1$ limit was
adopted to match that of the DEEP2 Survey and allows the ACES dataset
to probe even farther down the galaxy luminosity function at $z < 1$.

To maximize the sampling density of the survey at $0.2 < z < 1$,
stellar sources were down-weighted in the target selection
process. Stars were identified according to the spectral
classification of \citet{wolf04}, which utilized template SED fits to
the $17$-band photometry of COMBO-17. All sources classified as
``Star'' or ``WDwarf'' by \citet{wolf04} were down-weighted in the
target selection. This included a total of $\sim \! 1000$ sources at
$R < 23$, of which we targeted $189$ obtaining secure redshifts for
$162$. As illustrated in Figure \ref{fig_sampling_rate}, these stellar
sources are only a significant contribution to the total $R$-band
number counts at bright $(R < 21.5)$ magnitudes; excluding this
population of stars from the accounting, ACES targets $> \! 80\%$ of
sources at $R < 23$.

\begin{figure*}[tb!]
\centering
\plottwo{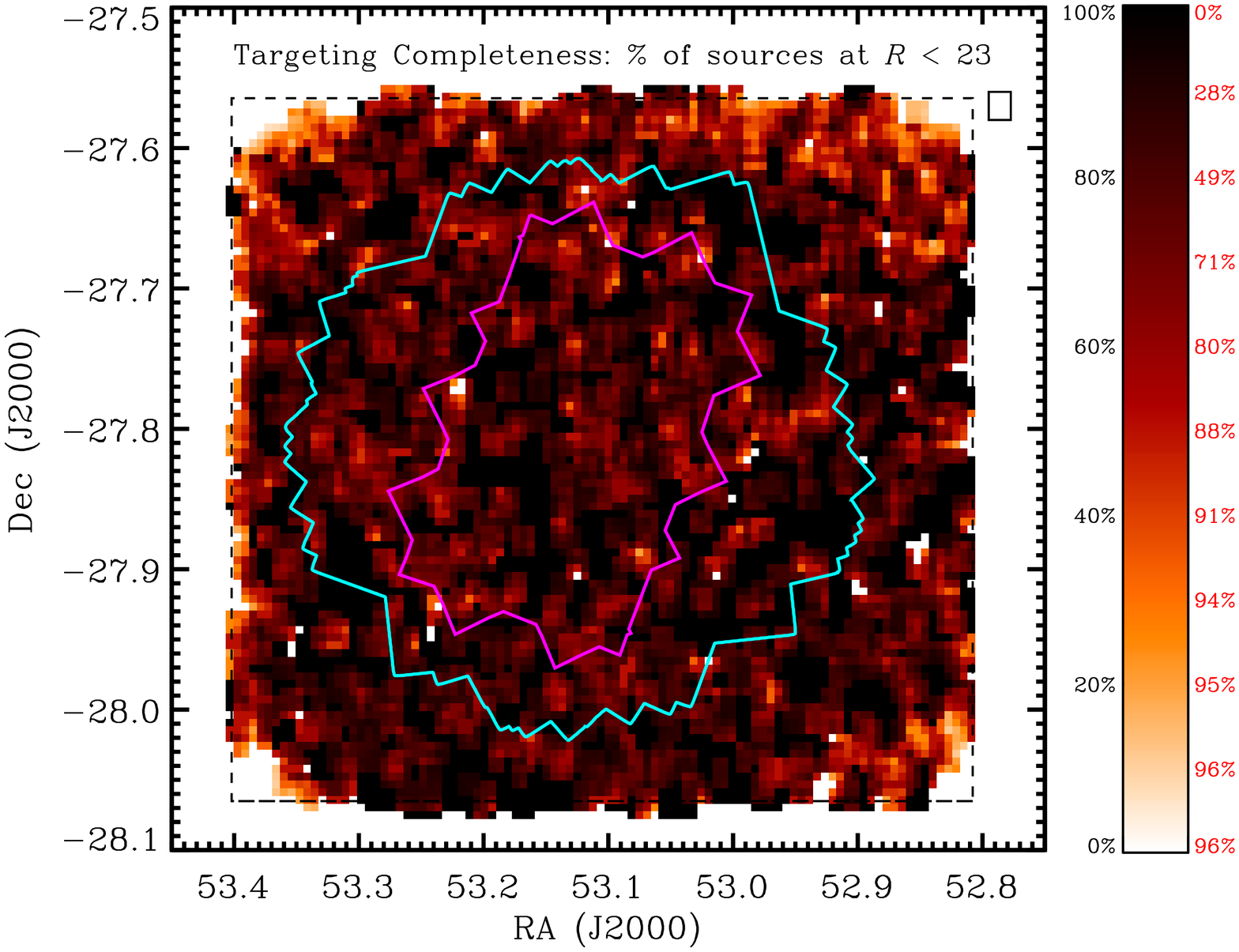}{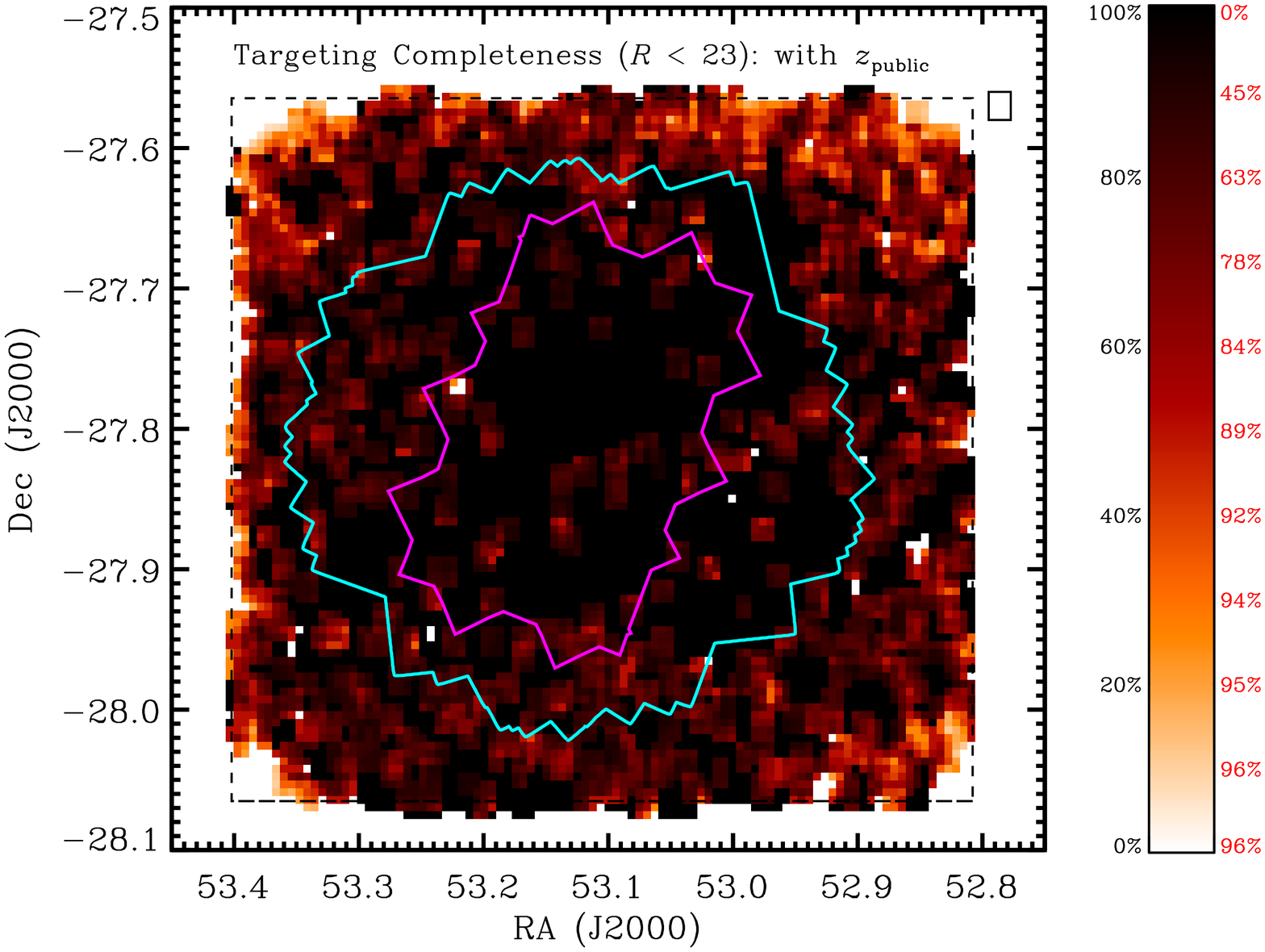}
\caption{The target sampling rate at $R_{\rm AB} < 23$ for the ACES
  target sample alone (\emph{left}) and for the joint population
  comprised of the ACES target sample and the set of existing public
  redshifts detailed in \S \ref{sec_design} (\emph{right}), computed
  in a sliding box of width $\Delta\alpha = 64^{\prime\prime}$ and
  height $\Delta\delta= 72^{\prime\prime}$. The size and shape of the
  box are illustrated in the upper right-hand corner of each plot. The
  associated color bars give the mapping from color to target
  completeness (where black and white correspond to $100\%$ and $0\%$
  completeness, respectively) and completeness is defined as the
  percentage of sources in the COMBO-17 imaging catalog with $R_{\rm
    AB} < 23$ (including stars) targeted by ACES (or ACES plus the set
  of sources with existing published redshifts). The red values to the
  right of each color bar show the portion of the $30^{\prime} \times
  30^{\prime}$ extended CDFS area (demarcated by the black dashed line
  in each plot) that has a target completeness greater than the
  corresponding level. Finally, the magenta and cyan outlines indicate
  the location of the GOODS {\it HST}/ACS and 2-Ms {\it
    Chandra}/ACIS-I observations, respectively. At $R < 23$, the
  sampling rate is exceptionally high ($\gtrsim \!  70\%$ from ACES
  alone) across nearly the entire extended CDFS. \vspace*{0.2in}}
\label{fig_sampling_map}
\end{figure*}

In selecting the ACES spectroscopic targets, we also down-weighted
many sources with an existing spectroscopic redshift in the
literature. This sample of ``public'' redshifts was drawn from
\citet{lefevre04}, \citet{vanzella05, vanzella06}, \citet{mignoli05},
\citet{ravikumar07}, \citet{szokoly04}, \citet{popesso09}, and
\citet{balestra10} as well as a small set of proprietary redshifts
measured by the DEEP2 Galaxy Redshift Survey using Keck/DEIMOS. In all
cases, only secure redshifts were employed. For example, only quality
``A'' and ``B'' (not ``C'') redshifts were included from
\citet{vanzella05, vanzella06, popesso09, balestra10}. At $R_{\rm AB}
< 24.1$, a total of $2149$ unique sources with a
spectroscopic redshift were down-weighted in the target selection
process, with $1218$ of these sources at $R < 23$. 
A significant number of objects ($1288$) in this public catalog were
observed by ACES; a comparison of the ACES redshift measurements to
those previously published is presented in \S \ref{sec_data}. Note
that some of these public redshifts (most notably those of
\citealt{balestra10}) were not yet published prior to the commencement
of ACES, but were included in the target selection process as the
survey progressed. Also, the distribution of the existing
spectroscopic redshifts on the sky is strongly biased towards the
center of the extended CDFS, primarily covering the smaller GOODS-S
region (see Fig.\ \ref{fig_sampling_map}).

\begin{figure}[h]
\centering
\plotone{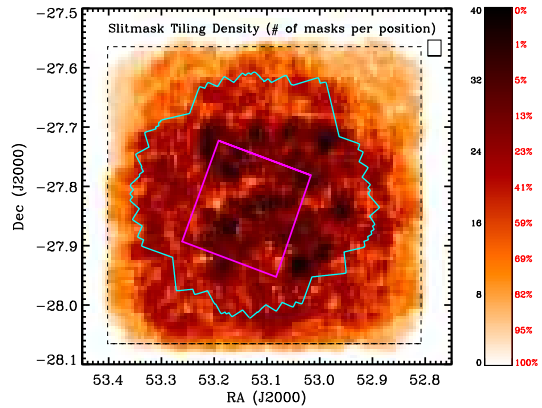}
\caption{The number of Magellan/IMACS slitmasks covering a given
  spatial location (computed within a box of width $\Delta\alpha =
  90^{\prime\prime}$ and height $\Delta\delta = 96^{\prime\prime}$) as
  a function of position within the CDFS. The red values to the right
  of the color bar show the portion of the $30^{\prime} \times
  30^{\prime}$ extended CDFS area (demarcated by the black dashed line
  in the plot) that is covered by greater than the corresponding
  number of slitmasks. ACES includes a total of $40$ slitmasks, with
  $82\%$ of the field covered by at least 8 slitmasks.  Finally, the
  magenta and cyan outlines indicate the location of the CANDELS {\it
    HST}/WFC3-IR and 2-Ms {\it Chandra}/ACIS-I observations,
  respectively. Every object in the field has multiple chances to be
  placed on an ACES slitmask, helping to achieve a high sampling
  density and minimize any bias against objects in overdense regions
  on the sky. \vspace*{0.2in}}

\label{fig_tiling}
\end{figure}

For the 2007B through 2009B observing seasons, slitmasks were designed
in pairs, sharing a common position and orientation. By placing two
masks at the same location on the sky, objects had multiple chances to
be included on a slitmask (and thus observed). Furthermore, we were
able to integrate longer on fainter targets by including those sources
on both of the masks at a given position, while only including
brighter sources on a single mask and thus maximizing the number of
sources observed. As discussed in \S \ref{sec_future}, data for
objects appearing on multiple slitmasks have yet to be combined; at
present, each slitmask is analyzed independently, such that there are
a higher number of repeated observations of some objects (especially
fainter targets).

The tiling scheme for the $40$ IMACS slitmasks was designed to cover
the extended $\sim \! 30^{\prime} \times 30^{\prime}$ area of the
extended CDFS, with the caveat that the position and orientation of
each slitmask was dictated by the availability of suitably bright
stars for guiding and dynamic focusing.\footnote{Note that the
  Magellan-Baade telescope includes an atmospheric dispersion
  corrector (ADC) for the f/11 Nasmyth position of IMACS.} A moderate
resolution grism and wide-band ($5650$--$9200$\AA) filter were
employed in the observations (see \S \ref{sec_redux}), allowing
multiple ($2$--$3$) targets to occupy a given spatial position on a
slitmask and thereby enable upwards of $\sim \! 400$ sources to be
observed on a single slitmask. The average number of targets per mask
was $\sim \!  350$, with the details of the slitmask design varying
slightly from year to year of observing.

For all of the ACES slitmasks, a standard $1^{\prime\prime}$ slitwidth
was employed, with a minimum slitlength of $\sim \! 7^{\prime\prime}$
(centered on the target) and a gap of at least $0.5^{\prime\prime}$
between slits. For a subset of the slitmasks, slits were extended to
fill otherwise unoccupied real estate on the slitmask. On every mask,
the set of possible targets was restricted to those sources for which
the entire spectral range of $5650$--$9200$\AA\ would fall on the
detector, given the position and orientation of the mask. The location
of the slitmasks on the sky was selected such that the full area of
the mask would fall within the CDFS region, leading to a slitmask
tiling scheme that more highly samples the central portion of the
field. This relative oversampling is a direct product of the large
size of the IMACS field-of-view at f/2 ($\sim \! 25^{\prime} \times
25^{\prime}$ unvignetted); any IMACS slitmask that falls entirely
within the extended CDFS region will overlap the central portion of
the CDFS, independent of orientation. As shown in Figure
\ref{fig_tiling}, the number of ACES slitmasks at a given position
within the CDFS varies significantly from $\gtrsim \! 30$ at the
center of the field to $\sim \! 10$ at the edges.

In spite of this spatial variation in the total sampling rate, ACES
achieves a relatively uniform spatial sampling rate for the main $(R <
23)$ galaxy sample. As evident in Figure \ref{fig_sampling_map}, ACES
targets $\gtrsim \! 80\%$ of sources at $R < 23$, relatively
independent of position within the CDFS. When including spectroscopic
observations from the literature (counting the associated objects as
being observed), the target sampling is remarkably complete at $R <
23$, with roughly half of the $\sim \! 30^{\prime} \times 30^{\prime}$
CDFS surveyed to $\sim \! 90\%$ completeness. This relatively high and
uniform sampling rate is critical for the ability to measure local
environments with the ACES dataset.

\section{Observations and Data Reduction}
\label{sec_redux}
ACES spectroscopic observations employed the f/2 camera in the
Inamori-Magellan Areal Camera and Spectrograph (IMACS) on the
Magellan-Baade telescope and were completed across four separate
observing seasons (2007B -- 2010B) as detailed in Table \ref{tab_obs}.
The instrument set-up included the 200 lines mm$^{-1}$ grism (blaze
angle of $15^{\circ}$) in conjunction with the WB5650--9200 wide-band
filter, which yields a nominal spectral coverage of $5650$--$9200$\AA\
at a resolution of $R \equiv \lambda / \Delta \lambda \sim 750$ (at
$7500$\AA). Each slitmask was observed for a total integration time of
$\sim \! 4500$--$7200$ sec, divided into (at least) $3$--$4$
individual $\sim \!  1500$--$1800$ sec integrations (with no dithering
performed) to facilitate the rejection of cosmic rays --- see Table
\ref{tab_obs} for details regarding the total integration
times. Immediately following each set of science exposures (i.e.,
without moving the telescope or modifying the instrument
configuration), a quartz flat-field frame and comparison arc spectrum
(using He, Ar, Ne) were taken to account for instrument flexure and
detector fringing.

There are two notable variations in the instrument configuration that
occurred in the course of the ACES observations. Between the 2007B
(January 2008) and 2008B (November 2008) observing seasons, the IMACS
CCDs were upgraded from the original SITe to deep-depletion E2V chips
\citep{dressler11}. The new CCDs provided much improved quantum
efficiency, especially at red wavelengths. In addition to the detector
update, for the initial observing run (in January 2008), the data were
collected using the incorrect grism. Instead of the 200 lines
mm$^{-1}$ grism, the higher-resolution 300 lines mm$^{-1}$ grism
(blaze angle of $17.5^{\circ}$) was installed in IMACS. The resulting
spectra from those slitmasks (ACES1--ACES8) are therefore at slightly
higher resolution ($R \equiv \lambda / \Delta \lambda \sim 1100$ at
$7500$\AA). Given that the slitmasks multiplex in the spectral (in
addition to the spatial) direction and were designed for use with the
lower-resolution grism, the spectra for many objects overlap
significantly. In addition, for a subset of the objects (primarily
those located closer to the edge of the slitmask), part of the
$5650$--$9200$\AA\ spectral window was dispersed off of the IMACS
detector. In such instances, on the order of $300$\AA\ was typically
lost at either the blue or red extreme of the spectral window. While
the data taken on this first observing run were negatively impacted by
the use of the incorrect grism (including a slight reduction in total
throughput), redshifts were still able to be measured for many of the
targets. In an attempt to maintain the uniformity of the dataset,
however, the vast majority of the objects on the ACES1--ACES8
slitmasks (1166 of 1315 objects) were reobserved in subsequent
observing seasons.

\begin{deluxetable*}{c c c c c c c c}
\tablewidth{0pt}
\tablecolumns{8}
\tablecaption{\label{tab_obs} Slitmask Observation Information}
\startdata
\hline
Slitmask & Observation Date & \multirow{2}{*}{$\alpha$
  (J2000)}\tablenotemark{{\it a}} &
  \multirow{2}{*}{$\delta$ (J2000)}\tablenotemark{{\it b}} &
  P.A.\tablenotemark{{\it c}} &
  \multirow{2}{*}{$N_{0}$}\tablenotemark{{\it d}} & 
  \multirow{2}{*}{$N_{z}$}\tablenotemark{{\it e}} & \multirow{2}{*}{Exposure
    Time}\tablenotemark{{\it f}} \\
Name & (UT) &  &  & (deg) &  & & \\
\hline \hline
ACES1 & 2008 Jan 02 & 03 32 22.000 & -27 53 25.00 & 90 & 376 & 160 & $7200$s \\
ACES2 & 2009 Jan 02 & 03 32 22.000 & -27 53 25.00 & 90 & 313 & 148 & $7200$s \\
ACES3 & 2008 Jan 04 & 03 32 16.610 & -27 43 58.29 & 90 & 327 & 168 & $7200$s \\
ACES4 & 2008 Jan 04 & 03 32 16.610 & -27 43 58.29 & 90 & 280 & 36 & $7200$s \\
ACES7 & 2008 Jan 03 & 03 32 33.983 & -27 54 00.00 & 90 & 339 & 137 & $7200$s \\
ACES8 & 2008 Jan 03 & 03 32 33.983 & -27 54 00.00 & 90 & 297 & 111 & $7200$s \\

ACES81 & 2008 Nov 25 & 03 32 33.750 & -27 53 48.58 & 90 & 303 & 170 & $7200$s \\
ACES82 & 2008 Nov 25 & 03 32 33.750 & -27 53 48.58 & 90 & 348 & 238 & $8631$s \\
ACES83 & 2008 Nov 25 & 03 32 16.500 & -27 54 45.00 & 90 & 343 & 194 & $7200$s \\
ACES84 & 2008 Nov 26 & 03 32 16.500 & -27 54 45.00 & 90 & 367 & 249 & $7200$s \\
ACES85 & 2008 Nov 26 & 03 32 40.000 & -27 48 30.00 & 0 & 359 & 273 & $7200$s \\
ACES86 & 2008 Nov 26 & 03 32 40.000 & -27 48 30.00 & 0 & 381 & 175 & $7400$s \\
ACES87 & 2008 Nov 27 & 03 32 25.690 & -27 49 40.00 & 90 & 376 & 202 & $7200$s \\
ACES88 & 2008 Nov 27 & 03 32 25.690 & -27 49 40.00 & 90 & 378 & 244 & $7200$s \\
ACES91 & 2008 Nov 28 & 03 32 51.000 & -27 47 45.00 & 0 & 366 & 212 & $7200$s \\
ACES92 & 2008 Nov 28 & 03 32 51.000 & -27 47 45.00 & 0 & 352 & 193 & $7200$s \\
ACES93 & 2008 Nov 27 & 03 31 57.000 & -27 47 45.00 & 180 & 364 & 206 & $7500$s \\
ACES94 & 2008 Nov 28 & 03 31 57.000 & -27 47 45.00 & 180 & 319 & 135 & $7740$s \\

ACES101 & 2009 Nov 14 & 03 32 25.690 & -27 49 40.00 & 90 & 361 & 172 & $7200$s \\
ACES102 & 2009 Nov 15 & 03 32 25.690 & -27 49 40.00 & 90 & 337 & 127 & $7200$s \\
ACES103 & 2009 Nov 14 & 03 32 33.750 & -27 53 48.58 & 90 & 345 & 143 & $9000$s \\
ACES104 & 2009 Nov 14 & 03 32 33.750 & -27 53 48.58 & 90 & 345 & 188 & $7200$s \\
ACES105 & 2009 Nov 16 & 03 32 16.500 & -27 54 45.00 & 90 & 339 & 132 & $7200$s \\
ACES106 & 2009 Nov 16 & 03 32 16.500 & -27 54 45.00 & 90 & 340 & 168 & $4800$s \\
ACES107 & 2009 Nov 15 & 03 32 40.000 & -27 48 30.00 & 0 & 348 & 116 & $6850$s \\
ACES108 & 2009 Nov 15 & 03 32 40.000 & -27 48 30.00 & 0 & 356 & 191 & $7200$s \\
ACES109 & 2009 Nov 16 & 03 31 57.000 & -27 47 45.00 & 180 & 340 & 127 & $5400$s \\
ACES110 & 2009 Nov 16 & 03 31 57.000 & -27 47 45.00 & 180 & 340 & 161 & $4500$s \\
ACES201 & 2010 Dec 09 & 03 32 30.500 & -27 49 50.00 & 0 & 370 & 216 & $4500$s \\
ACES202 & 2010 Dec 10 & 03 32 33.750 & -27 53 48.58 & 90 & 360 & 210 & $5400$s \\
ACES203 & 2010 Dec 10 & 03 31 57.000 & -27 47 45.00 & 180 & 362 & 243 & $6900$s \\
ACES204 & 2010 Dec 09 & 03 32 40.000 & -27 48 30.00 & 0 & 364 & 213 & $4500$s \\
ACES205 & 2010 Dec 09 & 03 32 16.500 & -27 54 45.00 & 90 & 352 & 111 & $4500$s \\
ACES206 & 2010 Dec 09 & 03 32 51.000 & -27 47 45.00 & 0 & 364 & 200 & $4500$s \\
ACES207 & 2010 Dec 11 & 03 32 30.000 & -27 48 00.00 & 90 & 368 & 227 & $5400$s \\
ACES208 & 2010 Dec 11 & 03 32 25.690 & -27 49 40.00 & 90 & 359 & 227 & $5400$s \\
ACES209 & 2010 Dec 10 & 03 32 25.000 & -27 46 15.00 & 90 & 361 & 242 & $5400$s \\
ACES210 & 2010 Dec 10 & 03 32 25.690 & -27 52 35.00 & 90 & 349 & 153 & $5100$s \\
ACES211 & 2010 Dec 11 & 03 32 10.000 & -27 48 00.00 & 0 & 360 & 210 & $5040$s \\
ACES212 & 2010 Dec 11 & 03 32 54.000 & -27 49 40.00 & 0 & 355 & 100 & $5400$s \\

\vspace*{-0.1in}

\enddata

\tablecomments{Details of the ACES Magellan/IMACS slitmasks.}

\tablenotetext{{\it a}}{Right ascension (in hr mn sc) of the slitmask
  center.}

\tablenotetext{{\it b}}{Declination (in deg min sec) of the slitmask
  center.}

\tablenotetext{{\it c}}{Position angle of the slitmask (E of N).}

\tablenotetext{{\it d}}{Number of targets on slitmask.}

\tablenotetext{{\it e}}{Number of secure $(Q=-1,3,4)$ redshifts measured on
  slitmask.}

\tablenotetext{{\it f}}{Total exposure time for slitmask (in seconds).}

\end{deluxetable*}

\begin{figure*}[tb!]
\centering
\plottwo{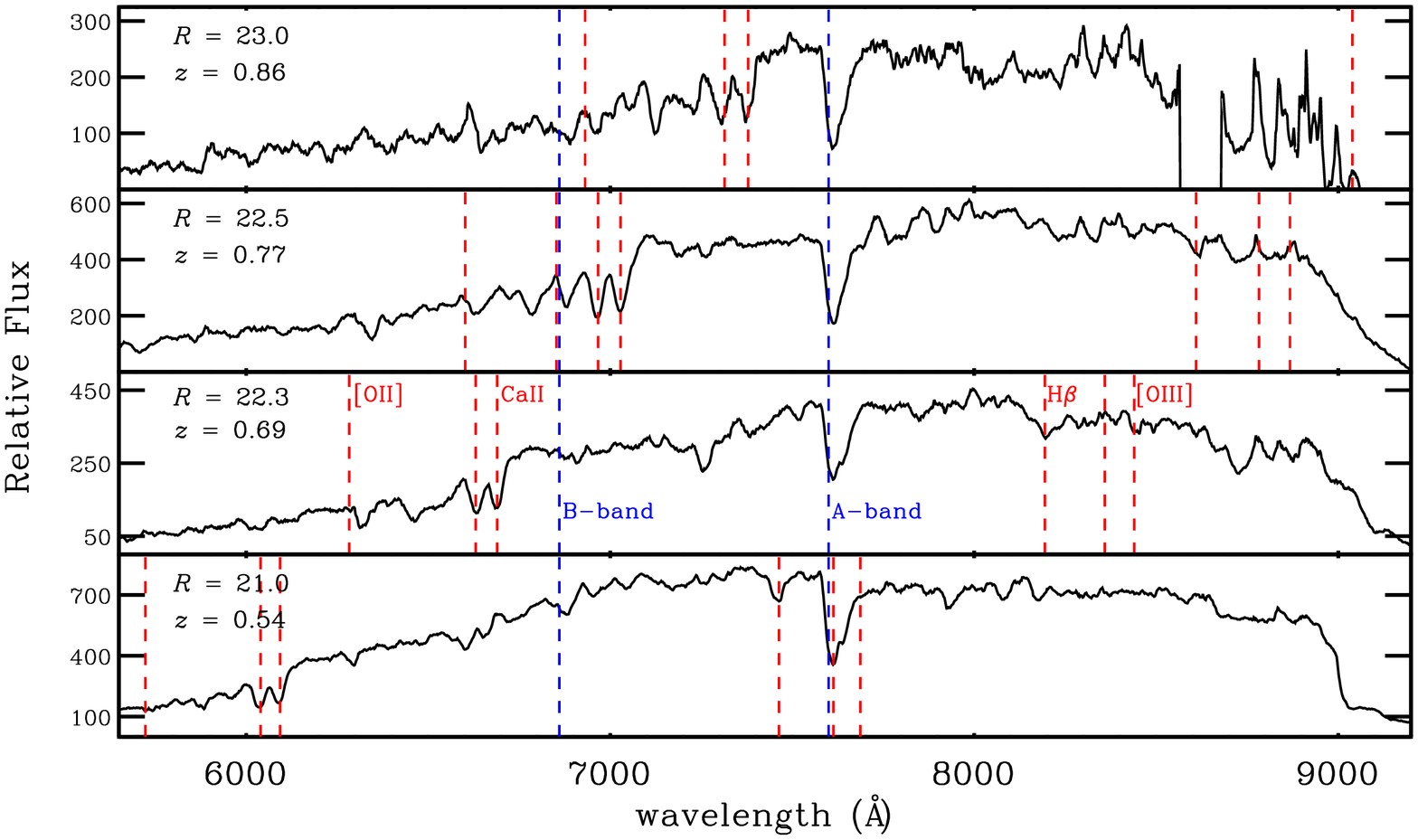}{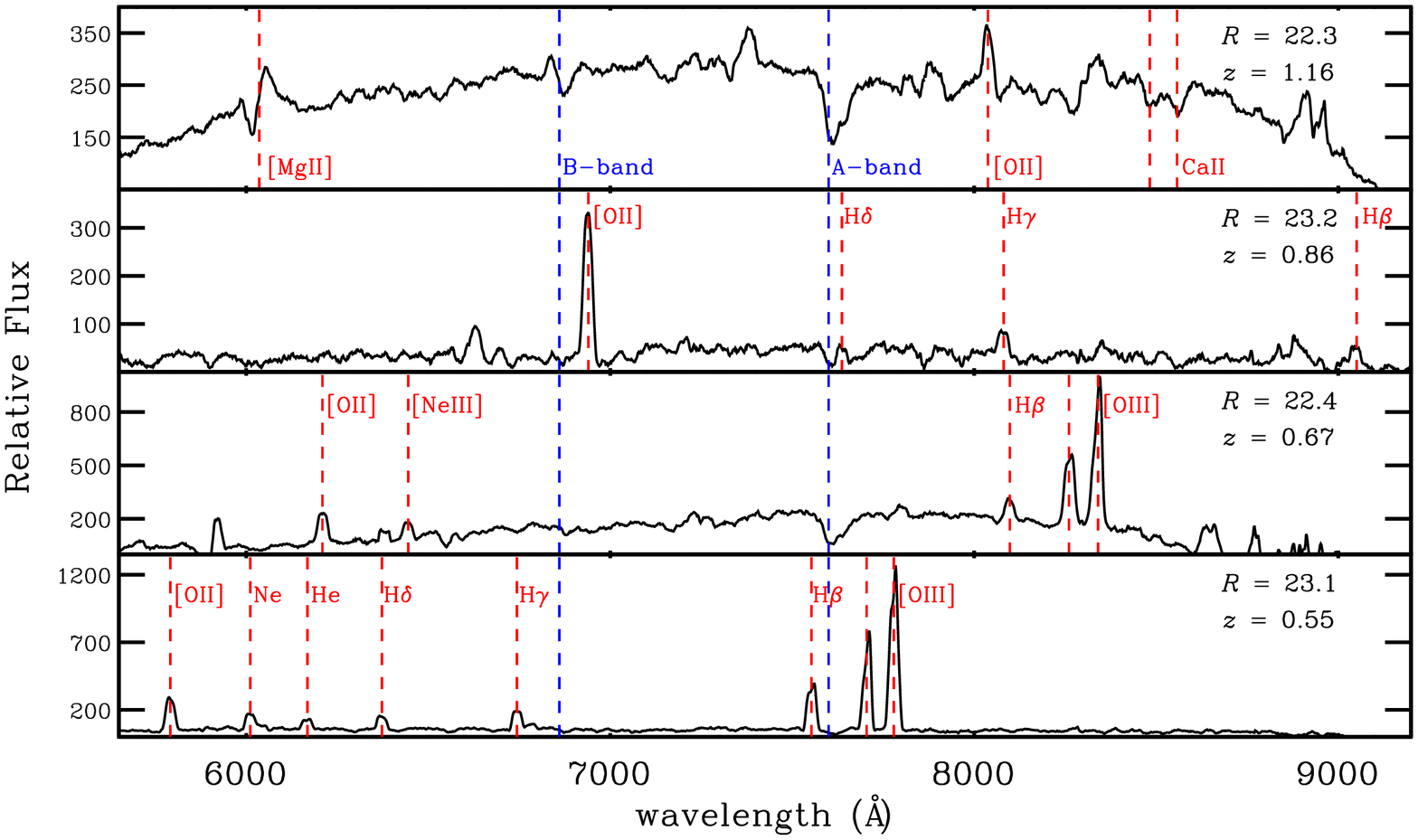}
\caption{Example ACES one-dimensional spectra of red and/or passive
  galaxies (\emph{left}) and star-forming/active galaxies
  (\emph{right}). The location of prominent spectral features as well
  as the A- and B-band telluric features are indicated by the red and
  blue dashed vertical lines, respectively. Note that the spectra have
  been smoothed (weighted by the inverse variance) using a kernel of
  $15$ pixels (or $30$\AA) in width. }
\label{fig_spectra}
\end{figure*}

The IMACS spectroscopic observations were reduced using the COSMOS
data reduction pipeline developed at the Carnegie Observatories
\citep{dressler11}.\footnote{http://obs.carnegiescience.edu/Code/cosmos}
For each slitlet, COSMOS yields a flat-fielded and sky-subtracted,
two-dimensional spectrum, with wavelength calibration performed by
fitting to the arc lamp emission lines. One-dimensional spectra were
extracted and redshifts were measured from the reduced spectra using
additional software developed as part of the DEEP2 Galaxy Redshift
Survey and adapted for use with IMACS as part of ACES and as part of
the spectroscopic follow-up of the Red-Sequence Cluster Survey (RCS,
\citealt{gladders05}; $z$RCS, Yan et al.\ in prep). A detailed
description of the DEEP2 reduction packages ({\it spec2d} and {\it
  spec1d}) is presented in \citet{cooper12b} and \citet{newman12}.

Example spectra for objects spanning a broad range of galaxy type and
apparent magnitude are shown in Figure \ref{fig_spectra}. All spectra
were visually inspected by M.~Cooper, with a quality code $(Q)$
assigned corresponding to the accuracy of the redshift value --- $Q =
-1,3,4$ denote secure redshifts, with $Q=-1$ corresponding to stellar
sources and $Q=3,4$ denoting secure galaxy redshifts (see Table
\ref{tab_catalog}). Confirmation of multiple spectral features was
generally required to assign a quality code of $Q = 3$ or $Q = 4$. As
discussed in \S \ref{public_comp}, $Q=3$ and $Q = 4$ redshifts are
estimated to be $> \! 90\%$ and $> \! 95\%$ reliable,
respectively. Quality codes of $Q=1,2$ are assigned to observations
that yield no useful redshift information ($Q=1$) or may possibly
yield redshift information after further analysis or re-reduction of
the data ($Q=2$). For detailed descriptions of the reduction pipeline,
redshift measurement code, and quality assignment process refer to
\citet{wirth04}, \citet{davis07}, and \citet{newman12}.

\section{Redshift Catalog}
\label{sec_data}

\begin{figure}[h!]
\centering
\plotone{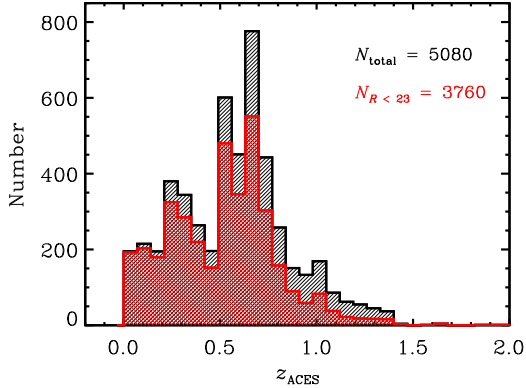}
\caption{The distribution of the $5080$ unique, secure $(Q = -1,3,4)$
  redshifts measured by ACES (black histogram). The red histogram
  shows the distribution for the main $(R < 23)$ target sample. The
  main sample is biased towards $z < 0.8$, with a tail to
  higher redshift.}
\label{fig_dndz}
\end{figure}

A preliminary ACES redshift catalog is presented in Table
\ref{tab_catalog}, a subset of which is listed herein. The entirety of
Table \ref{tab_catalog} appears in the electronic version of the
Journal. Note that a redshift is only included when classified as
being secure, $(Q=-1,3,4)$. The total number of secure redshifts in
the sample is $5080$ out of $7277$ total, unique targets. The redshift
distribution for this sample, as shown in Figure \ref{fig_dndz}, is
biased towards $z < 1$ with a tail out to higher redshift.

\begin{figure}
\centering
\plotone{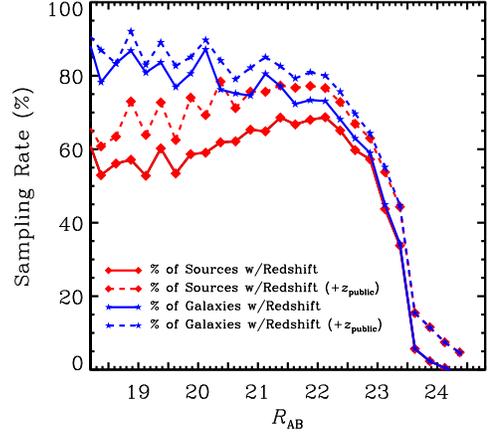}
\caption{The ACES redshift success rate as a function of $R$-band
  magnitude computed for all sources within the $\sim \! 30^{\prime}
  \times 30^{\prime}$ COMBO-17/CDFS footprint (solid red line). The
  redshift success rate is defined as the percentage of objects at a
  given $R$-band magnitude in the COMBO-17 photometric catalog
  (including stars) that were observed by ACES and yielded a secure
  ($Q = -1$,$3$,$4$ -- see \S \ref{sec_data}) redshift. The solid blue
  line shows the redshift completeness when only considering sources
  classified as non-stellar in the COMBO-17 catalog. At bright
  magnitudes, $R_{\rm AB} < 23$, the ACES sample is highly
  complete. The corresponding red and blue dashed lines show the
  associated completeness when accounting for sources with a published
  spectroscopic redshift in the literature (see \S \ref{sec_design}).}
\label{fig_zcompl_rate}
\end{figure}

At bright magnitudes ($R_{\rm AB} < 23$), ACES is highy complete,
obtaining a secure redshift for $\gtrsim \! 60\%$ of all sources
within the $\sim \! 30^{\prime} \times 30^{\prime}$ COMBO-17/CDFS
footprint (see Fig.\ \ref{fig_zcompl_rate}). When excluding sources
identified as stars by COMBO-17 ($\sim \! 1000$ sources; see \S
\ref{sec_design}) and including published spectroscopic redshifts from
the literature, the redshift completeness exceeds $80\%$ at the very
brightest magnitudes ($R_{\rm AB} < 22$).

\begin{figure}[h!]
\centering
\plotone{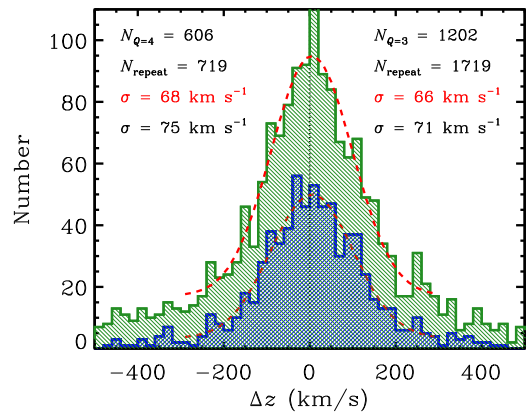}
\caption{The distribution of velocity differences computed from
  repeated observations, where both observations of a given galaxy
  yielded a $Q=4$ (blue histogram) or $Q=3$ redshift (green
  histogram). A total of $606$ and $1202$ unique sources comprise the
  two distributions separately. The dispersion as given by a
  Gaussian-fit to the distribution (red dashed line) and by the square
  root of the second moment of the distribution are reported in red
  and black font (or top and bottom numbers), respectively. A total of
  $2438$ pairs of observations comprise the two distributions, with a
  dispersion of $\sigma \sim 75$ km s$^{-1}$ independent of redshift
  quality. }
\label{fig_deltaz}
\end{figure}

As discussed in \S \ref{sec_design}, the ACES catalog has a high
number of repeated observations. These independent observations
provide a direct means for determining the precision of the redshift
measurements. As shown in Figure \ref{fig_deltaz}, we find a scatter
of $\sigma_{z} \sim 75$ km s$^{-1}$ within the ACES sample when
comparing repeat observations of a sizable sample of secure
redshifts. The scatter is found to be independent of the redshift
quality (i.e., $Q = 3$ versus $Q = 4$). While the precision of the
ACES redshifts is poorer than that of surveys such as DEEP2 and DEEP3,
it is adequate for characterizing local environments \citep{cooper05}
as well as identifying and measuring the velocity dispersions of
galaxy groups.

\subsection{Comparison to Photometric Redshift Samples}
\label{combo_comp}

The ACES spectroscopic sample provides an excellent dataset with which
to test the precision of the COMBO-17 photometric redshift
measurements. The COMBO-17 photo-$z$ estimates are based on $17$-band
photometry spanning 3500\AA\ to 9300\AA. While their accuracy at
higher redshift is impacted significantly by the lack of near-IR
observations, the COMBO-17 photometric redshifts are very robust at $z
< 1$. Based on a comparison to a relatively small ($< \! 1000$,
primarily at $z < 0.3$) sample of spectroscopic redshifts,
\citet{wolf04} found a $1$-$\sigma$ error of $\sigma_{z} / (1+z) < \!
0.01$, with a less than $1\%$ catastrophic failure rate (where failure
is defined to be $\Delta z / (1+z) > 0.05$).

\begin{figure}[h]
\centering
\plotone{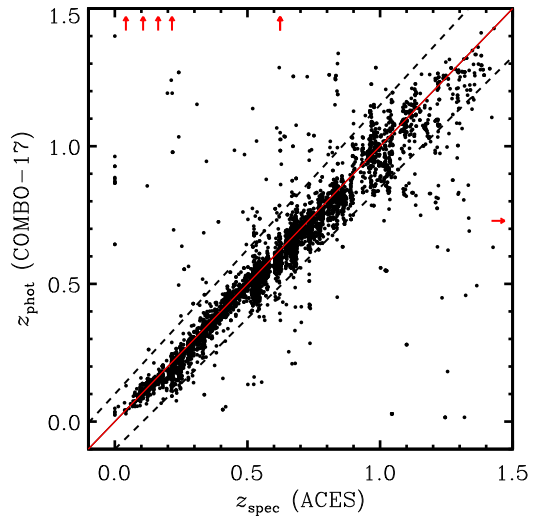}
\caption{A comparison of the spectroscopic redshifts from ACES to the
  photometric redshifts of COMBO-17 \citep{wolf04}. In general, the
  agreement is quite good, with a dispersion of $\sigma_{z} / (1+z)
  \sim 0.015$ (with $3$-$\sigma$ outliers removed) for the galaxy
  ($Q=3$,$4$) sample. The red vertical arrows indicate sources for
  which the photometric redshift value is greater than $z = 1.5$, in
  conflict with the spectroscopic value, which is indicated by the
  position of the arrow (and vice versa for the red horizontal
  arrow). The dashed black lines correspond to a nominal catastrophic
  failure level of $\Delta z / (1+z)= 0.05$.}
\label{fig_photoz}
\end{figure}

In Figure \ref{fig_photoz}, we directly compare the COMBO-17
photometric redshifts to the ACES spectroscopic redshifts for all
sources with a secure ($Q = -1$,$3$,$4$) redshift. The deficiency of
objects at $z \sim 0.9$ results largely from the inability of ACES to
resolve the [O {\footnotesize II}] doublet. That is, the ACES spectrum
of an emission-line galaxy at $z \sim 0.9$ would yield an unresolved
[O {\footnotesize II}] emission doublet at $\sim \! 7100$\AA, while
H$\beta$ and [O {\footnotesize III}] would be redward of our spectral
window ($> \! 9200$\AA). We are unable to easily distinguish this
single emission line from H$\alpha$ (i.e., a galaxy at $z \sim 0.08$),
as at that redshift H$\beta$ and [O {\footnotesize III}] would be
blueward of our spectral window ($< \! 5650$\AA). As a result, many
objects at $z \sim 0.9$ are classified as $Q=2$. Using broad-band
color info, we hope to recover these objects in the future
\citep[e.g.,][]{kirby07}.

Within the ACES dataset alone, there are $4769$ objects with a secure
galaxy redshift (i.e., $Q = 3$,$4$) and a photometric redshift in the
catalog of \citet{wolf04}. For this set of objects, the COMBO-17
photometric redshifts exhibit a dispersion of $\sigma_{z} / (1+z) \sim
0.015$ (with $3$-$\sigma$ outliers removed) and a catastrophic failure
rate (again taken to be $\Delta z / (1+z) > 0.05$) of $> \! 10\%$.
As highlighted by \citet{wolf04}, however, the COMBO-17 photometric
redshifts degrade in quality for increasingly fainter galaxies, and
the ACES sample extends to $R = 24.1$. For bright objects ($R \le
22$), the dispersion relative to the ACES spectroscopic redshifts is
$\sigma_{z} / (1+z) \sim 0.012$ (again with $3$-$\sigma$ outliers
removed), with a catastrophic failure rate of $6\%$. The precision is
slightly poorer at fainter magnitudes ($22 < R < 23$), increasing to
$\sigma_{z} / (1+z) \sim 0.014$, while at the faintest magnitudes
probed by ACES, the scatter between the COMBO-17 photo-$z$ values and
our spectroscopic redshifts increases to $\sigma_{z} / (1+z) \sim
0.022$ (for $R \ge 23$). For the main $R < 23$ sample, the
catastrophic failure rate ($\Delta z / (1+z) > 0.05$) is
$8\%$. 

\begin{figure*}[tb]
\centering
\plotone{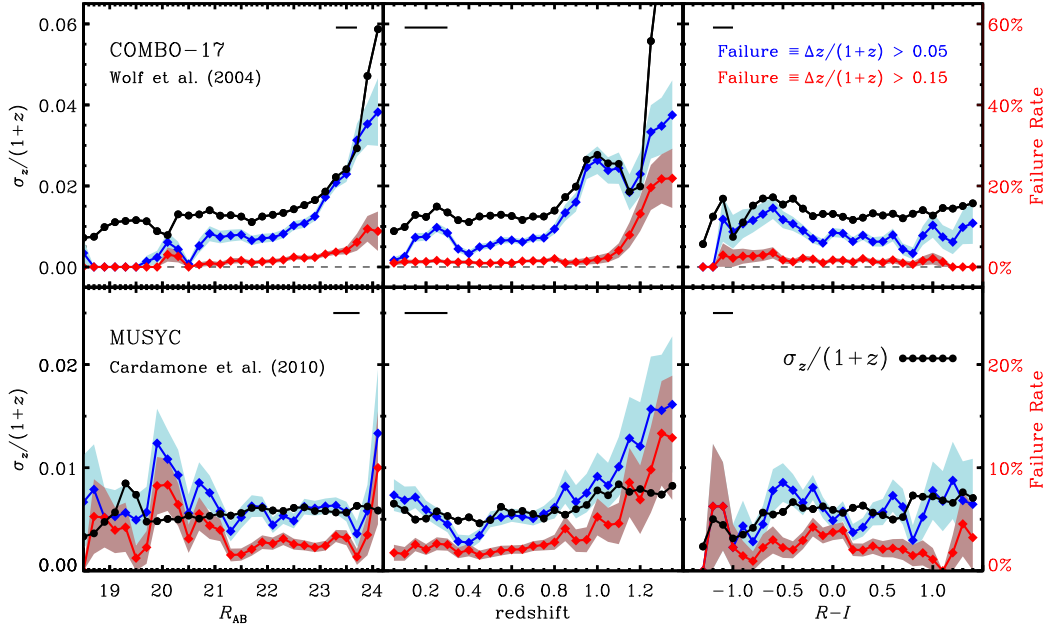}
\caption{The dependence of the photometric redshift error $(\sigma_{z}
  / (1+z))$ and the catastrophic failure rate on $R$-band magnitude
  (\emph{left}), redshift (\emph{middle}), and observed $R-I$ color
  (\emph{right}) for the COMBO-17 (\emph{top}) and MUSYC
  (\emph{bottom}) photometric redshift catalogs. The errors (black
  points) and failure rates (blue and red diamonds) are computed using
  sliding boxes with widths given by the black dashes in the upper
  corner of each plot, while the light blue and red shaded regions
  denote the $1\sigma$ uncertainty on the respective failure rates, as
  given by binomial statistics. In all cases, the dispersions
  $(\sigma_{z} / (1+z))$ are computed with $3\sigma$ outliers
  removed. In determining the redshift and color dependences, only
  objects with $R < 23$ are included. For COMBO-17, the photo-$z$
  errors and failure rates increase significantly at fainter
  magnitudes $(R > 23)$ and higher redshift $(z > 1)$, while the
  trends are much weaker for the MUSYC photometric
  redshifts. \vspace*{0.2in}}
\label{fig_photoz_err}
\end{figure*}

These trends with apparent magnitude and redshift are evident in
Figure \ref{fig_photoz_err}, which shows the dependence of the
photometric redshift error $(\sigma_{z} / (1+z))$ and the catastrophic
failure rate on $R$-band magnitude, redshift, and observed $R-I$ color
based on a comparison of the ACES spectroscopic redshift and COMBO-17
photometric redshift catalogs. At faint magnitudes $(R > 23)$ and at
higher redshift $(z > 1)$, the photometric-redshift errors and failure
rates for COMBO-17 increase significantly. However, we find no
significant correlation between the quality of the photometric
redshifts and apparent $R-I$ color, suggesting that there is little
dependence on the spectral-type or star-formation history of a galaxy.

As highlighted earlier, the degradation in photo-$z$ quality with
redshift is in part due to the lack of near-IR photometry in the
multi-band imaging of COMBO-17. In contrast, the photometric redshifts
from the Multiwavelength Survey by Yale-Chile
\citep[MUSYC,][]{gawiser06}, as computed by \citet{cardamone10},
include broad-band optical and near-IR ($JHK$) imaging in addition to
photometry in $18$ medium-bands from Subaru. As shown in Figure
\ref{fig_photoz_err}, the MUSYC photometric redshifts exhibit much
smaller scatter in relation to the ACES spectroscopic redshift sample,
with $\sigma_{z} / (1+z) \sim 0.005$ across the full magnitude and
redshift range probed. In addition, the catastrophic failure rate for
the MUSYC sample is roughly a factor of $2$ lower than that found for
the COMBO-17 photometric redshift catalog.

\subsection{Comparison to Spectroscopic Redshift Samples}
\label{public_comp}

\begin{figure}[h!]
\centering
\plotone{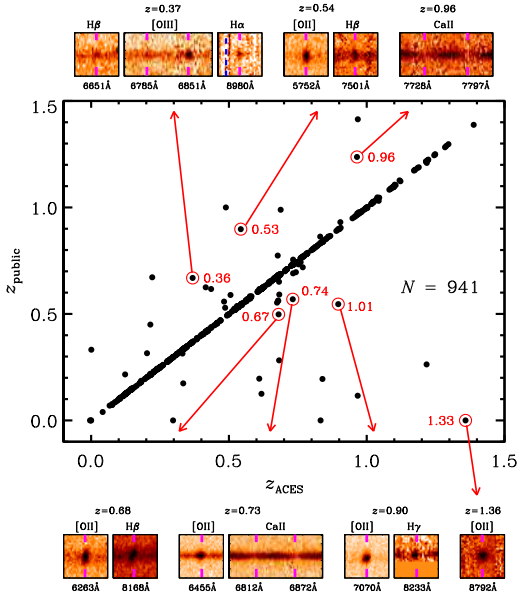}
\caption{For the sample of $941$ objects with a secure redshift in
  ACES (i.e., $Q = -1$, $3$, or $4$) and also a secure measurement in
  the literature, we plot a comparison of the two spectroscopic
  redshift measurements. The agreement is quite good, with only $47$
  objects having redshift measurements that disagree at $|\Delta z| >
  3000$ km s$^{-1}$. For $7$ of these outliers, we show cut-outs from
  the Magellan/IMACS two-dimensional spectra, illustrating the
  spectral features that confirm the ACES redshift. In each case, the
  COMBO-17 photometric redshift (given in red font within the plot)
  agrees quite well with the ACES spectroscopic redshift. Refer to \S
  \ref{sec_data} for details regarding the set of ``public'' redshift
  measurements. }
\label{fig_speczcomp}
\end{figure}

Matching our catalog to previously-published spectroscopic redshifts
in the field \citep[e.g.,][see \S \ref{sec_design}]{lefevre04,
  vanzella05, vanzella06, balestra10}, we find $1288$ of our targets
have a redshift published as part of these existing datasets. For
$941$ of these $1288$ objects, we measure a secure redshift from our
IMACS spectroscopy. The agreement between the ACES redshifts and those
in the literature is generally good. We find a median offset of
$|\Delta z| \sim \! 240$ km s$^{-1}$ when comparing to the ``public''
redshift catalog detailed in \S \ref{sec_design}. For the small set of
significant outliers (the $44$ objects with $|\Delta z| > 3000$ km
s$^{-1}$), the ACES spectra were re-examined to confirm the validity
of the ACES redshifts. While some outliers could be the result of
mismatching between the ACES catalog and the public databases, the
majority are the result of line misidentification (e.g., confusing
H$\alpha$ with [O {\footnotesize II}]) or some other failure in
redshift identification (see Figure \ref{fig_speczcomp}).

The significant overlap between the ACES sample and the set of
existing redshifts in the literature also provides a means to
conservatively estimate the reliability of the ACES redshifts. Taking
the previously-published values to be the true redshift for each
galaxy, we measure the catastrophic failure rate ($|\Delta z| =
|z_{\rm public} - z_{\rm ACES}| > 1000$ km s$^{-1}$) for the $Q=3$ and
$Q=4$ ACES redshifts. For the $351$ sources with $Q=3$ and $586$
source with $Q=4$ redshifts in the ACES catalog, we find failure rates
of $13\%$ and $6\%$, respectively. As shown in Figure
\ref{fig_speczcomp}, some of the previously-published redshifts are
clearly in error, thus these confidence values are conservative
estimates. Comparing within the ACES sample alone, we find that $6\%$
and $2\%$ of sources with repeated observations (both yielding $Q=3$
and $Q=4$ redshifts --- see Fig.\ \ref{fig_deltaz}) have redshifts
measurements that disagree at greater than $500$ km s$^{-1}$.

The new redshifts presented here should significantly enhance studies
of galaxy evolution and cosmology in the CDFS. Our sample expands upon
previous spectroscopic work in the field, significantly increasing the
size of the existing redshift database. Furthermore, our observations
broaden the area covered, extending beyond the GOODS-S {\it HST}/ACS
footprint, allowing us to target a greater number of relatively rare
sources. In particular, we specifically targeted {\it Spitzer}/MIPS
$70\mu$m sources, including those observed by previous spectroscopic
efforts in the field. Within the FIDEL Survey's {\it Spitzer}/MIPS
$70\mu$m photometric catalog for GOODS-S, which covers an area of
roughly $10^{\prime} \times 10^{\prime}$, there are only $44$ sources
detected at $> \! 2.5$ mJy \citep{magnelli11}. The relatively small
number of these sources puts a premium on spectroscopic follow-up,
including those located outside of the GOODS-S area. The FIDEL Survey
covers a broader region surrounding the GOODS-S area, actually
extending significantly beyond the ACES footprint in most directions
when combined with existing {\it Spitzer}/MIPS observations. In total,
$\gtrsim \!  500$ sources are detected (down to $S_{70\mu{\rm m}} \sim
0.5$ mJy) at $70\mu$m within the COMBO-17 footprint as part of the
FIDEL survey (requiring a $3$-$\sigma$ detection at both $24\mu$m and
$70\mu$m). As highlighted in \S \ref{sec_design}, ACES targets $529$
sources as potential optical counterparts to these sources (i.e.,
within $3^{\prime\prime}$ of a $70\mu$m source).

The $70\mu$m observations conducted as part of the FIDEL Survey are
the deepest in the sky, allowing significant numbers of star-forming
galaxies and active galactic nuclei to be detected out to intermediate
redshift at rest-frame wavelengths that are dramatically less impacted
by aromatic and silicate emission than those normally probed by {\it
  Spitzer}/MIPS $24\mu$m observations. With accompanying redshift
information from spectroscopic follow-up such as presented here, these
deep far-infrared data provide a unique constraint on the cosmic
star-formation history at intermediate redshift
\citep[e.g.,][]{magnelli09}.

\begin{figure*}[tb!]
\centering
\plottwo{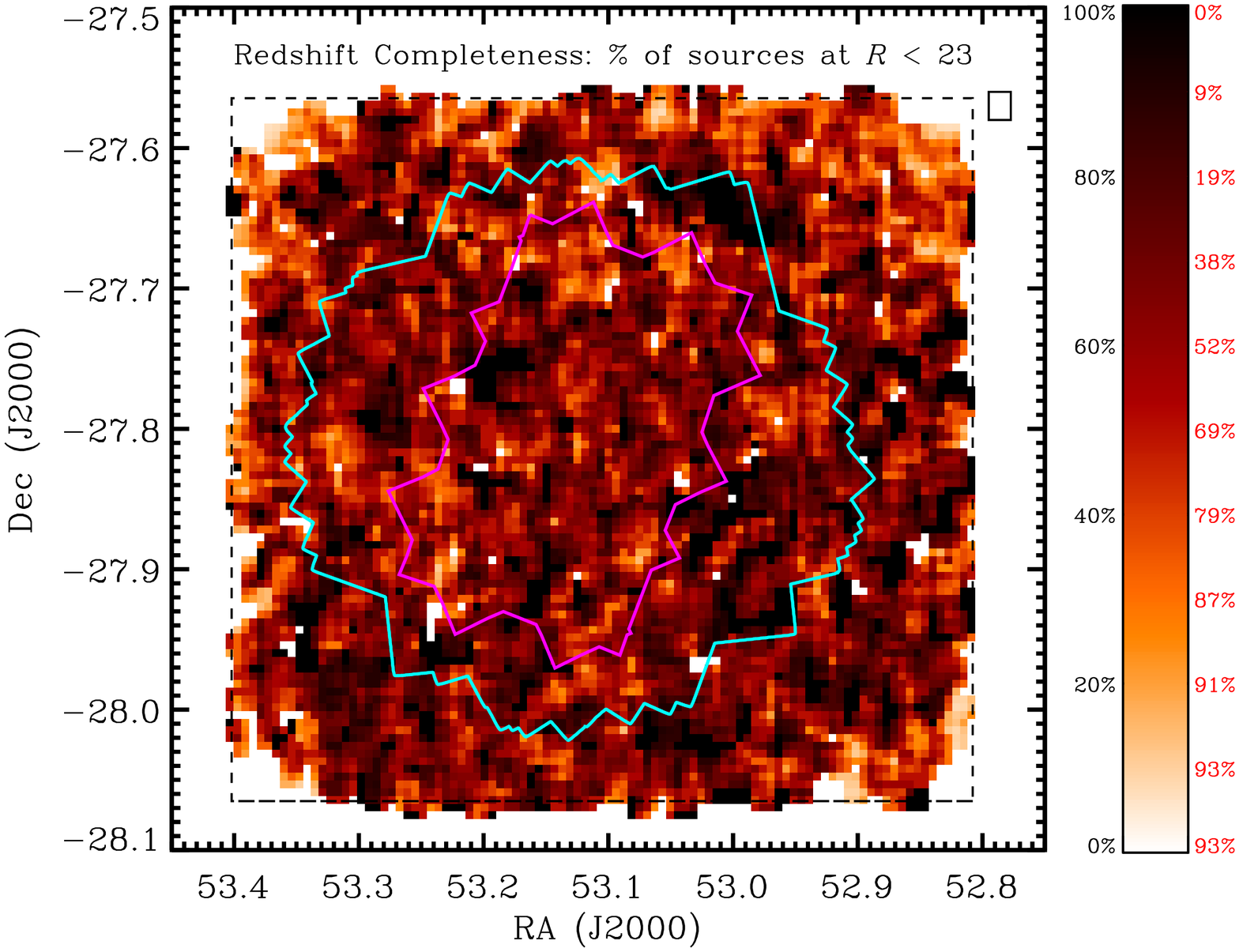}{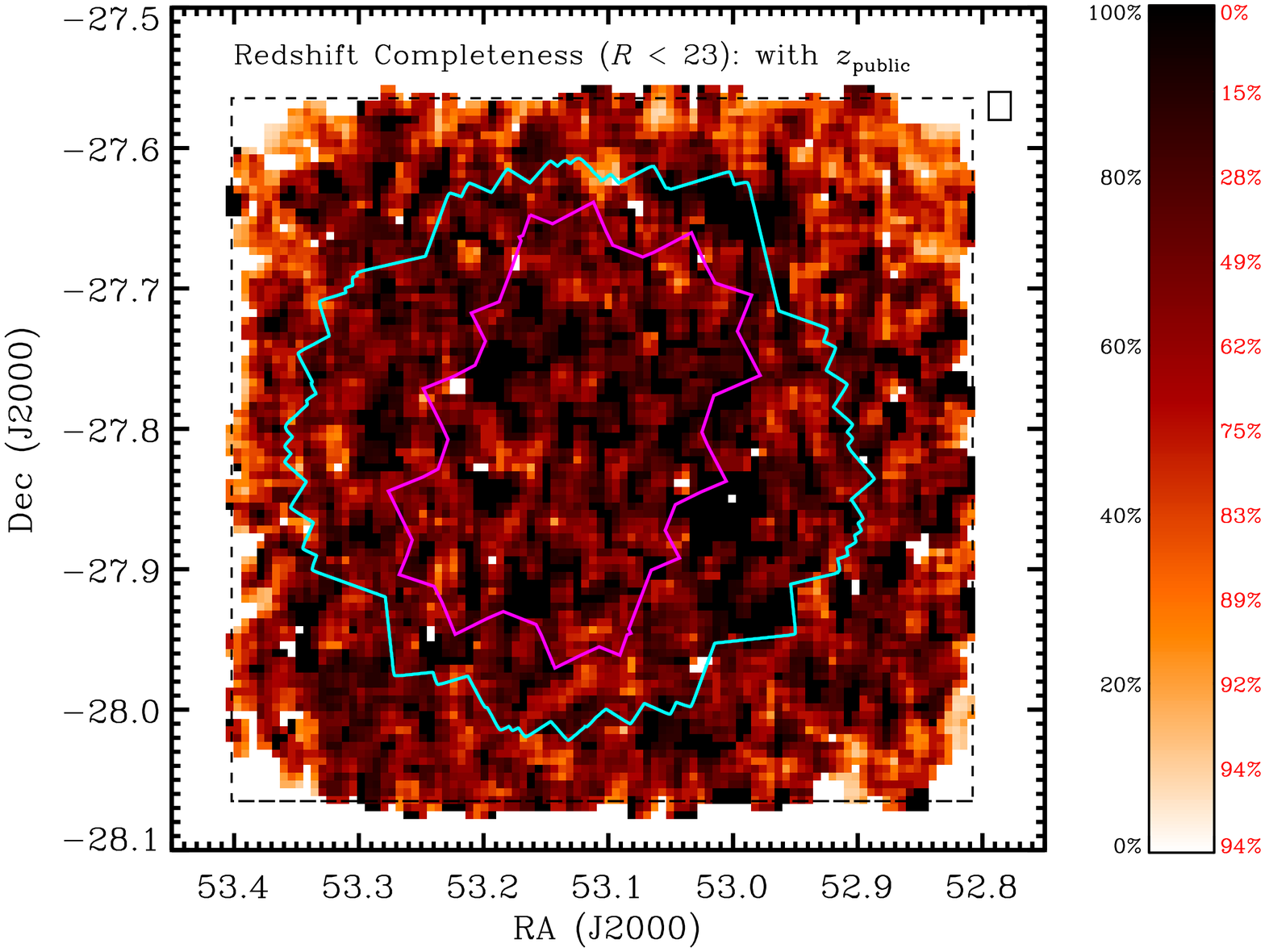}
\caption{The redshift completeness at $R_{\rm AB} < 23$ for the ACES
  sample alone (\emph{left}) and for the joint population comprised by
  ACES and the set of existing public redshifts detailed in \S
  \ref{sec_design} (\emph{right}), computed in a sliding box of width
  $\Delta\alpha= 64^{\prime\prime}$ and height $\Delta\delta =
  72^{\prime\prime}$. The size and shape of the box are illustrated in
  the upper right-hand corner of each plot. The associated color bars
  give the mapping from color to redshift completeness (where black
  and white correspond to $100\%$ and $0\%$ completeness,
  respectively) and completeness is defined as the percentage of
  sources in the COMBO-17 imaging catalog with $R_{\rm AB} < 23$
  (including stars) for which ACES (or ACES plus the set of sources
  with existing published redshifts) measured a secure redshift (i.e.,
  $Q = -1,3,4$). The red values to the right of each color bar show
  the portion of the $30^{\prime} \times 30^{\prime}$ extended CDFS
  area (demarcated by the black dashed line in each plot) that has a
  redshift completeness greater than the corresponding level. Finally,
  the magenta and cyan outlines denote the location of the GOODS
  {\it HST}/ACS and 2-Ms {\it Chandra}/ACIS-I observations,
  respectively. At $R < 23$, the redshift completeness is high
  ($\gtrsim \! 50\%$ from ACES alone) across nearly the entire
  extended CDFS. \vspace*{0.4in}}
\label{fig_fullzcompl}
\end{figure*}

\begin{deluxetable*}{c c c c c c c c c c c c c c}
\tablewidth{0pt}
\tablecolumns{14}
\tablecaption{\label{tab_catalog} ACES Redshift Catalog}
\tablehead{Object ID\tablenotemark{{\it a}} & $\alpha$\tablenotemark{{\it b}}
  (J2000) & $\delta$\tablenotemark{{\it c}} (J2000) & $R_{\rm
    AB}$\tablenotemark{{\it d}} & Mask\tablenotemark{{\it e}} &
  Slit\tablenotemark{{\it f}} & MJD\tablenotemark{{\it g}} &
  flag$_{70\mu{\rm m}}$\tablenotemark{{\it h}} & $z$\tablenotemark{{\it i}} &
  $z_{\rm helio}$\tablenotemark{{\it j}} & $Q$\tablenotemark{{\it k}} &
  $z_{\rm  other}$\tablenotemark{$\ell$} & Ref\tablenotemark{{\it m}}} 
\startdata
122 & 53.038740 & -28.064493 & 22.54 & ACES105 & 152 & 55150.9 &
 0 & 0.38687 & 0.38686 & 4 & \ldots & \ldots \\ 
215 & 53.053105 & -28.063773 & 21.05 & ACES106 & 163 & 55151.7 & 
 0 & \ldots & \ldots & 1 & \ldots & \ldots \\
15629 & 52.978946 & -27.943812 & 23.15 & ACES102 & 085 & 55150.6 & 
 0 & 0.74274 & 0.74272 & 3 & \ldots & \ldots \\
\vspace*{-0.1in}
\enddata
\tablecomments{Table \ref{tab_catalog} is presented in its entirety in
  the electronic edition of the Journal. A portion is shown here for
  guidance regarding its form and content.}

\tablenotetext{{\it a}}{Object identification number (SEQ) in $R$-band
  catalog of \citet{wolf04}.}

\tablenotetext{{\it b}}{Right ascension in decimal degrees from
  \citet{wolf04}.}

\tablenotetext{{\it c}}{Declination in decimal degrees from
  \citet{wolf04}.}

\tablenotetext{{\it d}}{$R$-band magnitude in AB system from
  \citet{wolf04}.}

\tablenotetext{{\it e}}{Name of IMACS slitmask on which object was
  observed.}

\tablenotetext{{\it f}}{Number of slit on IMACS slitmask corresponding to
  object.}

\tablenotetext{{\it g}}{Modified Julian Date of observation.}

\tablenotetext{{\it h}}{Targeting flag: 0 = main $R$-band selected target; 1
  = {\it Spitzer}/MIPS $70\mu$m target}

\tablenotetext{{\it i}}{Redshift derived from observed spectrum.}

\tablenotetext{{\it j}}{Heliocentric-frame redshift.}

\tablenotetext{{\it k}}{Redshift quality code (${\rm star} = -1$; $\sim \!
  90\%\ {\rm confidence} = 3$; $\sim \! 95\%\ {\rm confidence} = 4$;
  ${\rm unknown} = 1,2$).}

\tablenotetext{$\ell$}{Alternate redshift from literature.}

\tablenotetext{{\it m}}{Source of alternate redshift: (1) DEEP2/DEIMOS ;
  (2) \citet{lefevre04}; (3) \citet{vanzella05}; (4)
  \citet{vanzella06}; (5) \citet{mignoli05}; (6) \citet{ravikumar07};
  (7) \citet{szokoly04}; (8) \citet{popesso09, balestra10}}

\end{deluxetable*}

\section{Environment Measures}
\label{sec_environ}

By extending beyond the GOODS-S footprint (i.e., the area primarily
targeted by previous spectroscopic efforts in the field), ACES
substantially expands the area over which galaxy overdensity (or
``environment'') can be measured in the CDFS. The finite area of sky
covered by a survey introduces geometric distortions --- or edge
effects --- which bias environment measures near borders (or holes) in
the survey field, generally leading to an underestimate of the local
overdensity \citep{cooper05, cooper06}. To minimize the impact of
these edge effects on studies of environment, galaxies near the edge
of the survey field (e.g., within a projected distance of $1$-$2\
h^{-1}$ comoving Mpc of an edge) are often excluded from any
analysis. As such, the ACES dataset, which spans a considerably larger
region than previous spectroscopic samples (and with a much more
spatially-uniform sampling rate, see Fig.\ \ref{fig_sampling_map} and
Fig.\ \ref{fig_fullzcompl}), now allows the environment of galaxies at
intermediate redshift to be accurately computed across nearly the
entire $\sim 30^{\prime} \times 30^{\prime}$ area of the CDFS, thereby
enabling unique analyses of small-scale clustering in one of the most
well-studied extragalactic fields in the sky.

For each galaxy in the ACES redshift catalog (see Table
\ref{tab_catalog}), we estimate the local galaxy overdensity, or
``environment'', using measurements of the projected
third-nearest-neighbor surface density ($\Sigma_{3}$) about each
galaxy, where the surface density depends on the projected distance to
the third-nearest neighbor, $D_{p,3}$, as $\Sigma_{3} = 3 / (\pi
D_{p,3}^{2})$. Over quasi-linear regimes, the mass density and galaxy
density should simply differ by a factor of the galaxy bias
\citep{kaiser87}. In computing $\Sigma_{3}$, only objects within a
velocity window of $\pm1250$ km s$^{-1}$ are counted, to exclude
foreground and background galaxies along the line-of-sight. To explore
any dependencies on the choice of $N$ in this
$N^{th}$-nearest-neighbor approach to measuring environment, we also
compute overdensities based on the distance to the fourth- and
fifth-nearest-neighbor (see Table \ref{tab_environ}).

When estimating the local environment within a survey dataset, each
surface density measurement must be corrected according to the
redshift and spatial dependence of the survey's sampling rate. To
minimize the variation in the spatial component of the ACES sampling
rate, we select the $R < 23$ galaxy population as the tracer
population by which the local galaxy density is defined --- note that
this is done both with and without the public redshifts included and
environment measures based on each tracer population are provided in
Table \ref{tab_environ}.\footnote{See \citet{cooper09, cooper10b} for
  additional discussion regarding the selection of tracer populations
  in the measurement of environments.} While selecting only those
objects that meet this bright magnitude limit decreases the sampling
density of the tracer population (relative to the full ACES dataset),
the main $R < 23$ galaxy sample has a well-defined and relatively
uniform spatial selection rate (see Fig.\ \ref{fig_sampling_map} and
Fig.\ \ref{fig_fullzcompl}). Using this highly-complete tracer
population, we measure the surface density, $\Sigma_{3}$ (as described
above), about all galaxies in the ACES redshift catalog, independent
of apparent magnitude. With or without the public redshifts included,
the typical projected distance to the third-nearest neighbor,
$D_{p,3}$, is $\sim \! 1$ $h^{-1}$ Mpc at $0.2 < z < 0.8$.

To account for the relatively modest variations in completeness across
the field, each surface density measure is divided by the redshift
completeness at $R_{\rm AB} < 23$ (computed within a window
corresponding to $1$ $h^{-1}$ comoving Mpc$^{2}$ centered on each
object). We define the size of the window, in this redshift-dependent
manner, to roughly correspond to the typical distance to the projected
$3^{\rm rd}$-nearest neighbor. The redshift completeness value is only
weakly dependent on the size of the window employed; for example, the
use of a window with fixed size (e.g., $\sim \! 60^{\prime\prime}$ or
$\sim \! 120^{\prime\prime}$ on a side) yields similar results. Due to
the high level of completeness achieved by ACES, the resulting
correction applied to the surface density measurements are remarkably
modest, with the resulting environment measures highly correlated to
those computed without correction for variations in redshift
completeness.\footnote{Estimating environment with and without
  applying corrections for redshift incompleteness yields overdensity
  measures, $\log_{10}(1+\delta_{N})$, that are highly correlated with
  Pearson and ranked Spearman correlation coefficients of $0.99$.}
Within the central portion of the CDFS field, the variation in
redshift completeness at $R_{\rm AB} < 23$ (including public
redshifts) is well fit by a Gaussian centered at $\sim \!  0.8$ (i.e.,
$80\%$ completeness) and with a dispersion of $\sigma < 0.1$.

To correct for the redshift dependence of the ACES sampling rate, each
surface density is divided by the median $\Sigma_{3}$ for all galaxies
within a window $\Delta z = 0.03$ centered on the redshift of each
galaxy; this converts the $\Sigma_{3}$ values into measures of
overdensity relative to the median density (given by the notation $1 +
\delta_{3}$ herein) and effectively accounts for the redshift
variations in the selection rate \citep{cooper05, cooper06,
  cooper08b}. We restrict our environment catalog to the redshift
range $0.2 < z < 0.8$, avoiding the low- and high-redshift tails of
the ACES redshift distribution (see Fig.\ \ref{fig_dndz}) where the
variations in the survey selection rate are the greatest.

Finally, to enable the effects of edges and holes in the survey
geometry to be minimized, we measure the distance to the nearest
survey boundary. We determine the survey area and corresponding edges
according to the 2-dimensional survey completeness map
($w(\alpha,\delta)$, see Fig.\ \ref{fig_fullzcompl}) and the
photometric bad-pixel mask, which provides information about the
location of bright stars (i.e., undersampled regions) in the field. We
define all regions of sky with $w(\alpha,\delta) < 0.3$ averaged over
scales of $\gtrsim \!  30^{\prime\prime}$ to be unobserved and reject
all significant regions of sky ($\gtrsim \!  30^{\prime\prime}$ in
scale) that are incomplete in the COMBO-17 $R$-band photometric
catalog. Areas of incompleteness on scales smaller than
$30^{\prime\prime}$ are comparable to the typical angular separation
of galaxies targeted by ACES and thus cause a negligible perturbation
to the measured densities. To minimize the impact of edges on the data
sample, we recommend all analyses using these environment values to
exclude any galaxy within $1$ $h^{-1}$ comoving Mpc of an edge or
hole; such a cut greatly reduces the portion of the dataset
contaminated by edge effects \citep{cooper05}.

\begin{figure}[h!]
\centering
\plotone{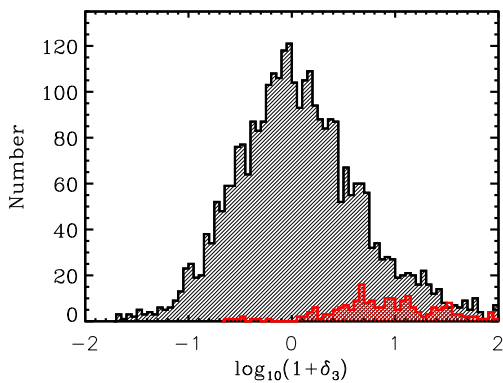}
\caption{The distribution of overdensity measures for all sources with
  a secure redshift at $0.2 < z < 0.8$ in the joint population comprised
  of the ACES redshift catalog and the set of existing public
  redshifts detailed in \S \ref{sec_design}. The red histogram shows
  the environment distribution for the $210$ galaxies identified as
  group members using the X-ray group catalog of Finoguenov et al.\
  (2011). There is good agreement between the group and environment
  catalogs. \vspace*{0.3in}}
\label{fig_environ}
\end{figure}

In Figure \ref{fig_environ}, we show the distribution of
overdensities, $\log_{10}(1+\delta_{3})$, for $3057$ galaxies with a
secure redshift at $0.2 < z < 0.8$ in either the ACES redshift catalog
or the set of existing public redshifts detailed in \S
\ref{sec_design}. Here, we exclude all galaxies within $1$ $h^{-1}$
comoving Mpc of a survey edge and utilize the environment measures
computed with a tracer population comprised of all galaxies at $R <
23$ (using both ACES and ''public'' secure redshifts). In addition,
Figure \ref{fig_environ} shows the distribution of environments for
those galaxies identified as members of X-ray groups by Finoguenov et
al.\ (2011). Group members are selected within a cylinder with a
radius of $0.75$ $h^{-1}$ Mpc and length of $2000$ km s$^{-1}$,
centered on the location of the extended X-ray emission for all groups
with $M_{200} > 5 \cdot 10^{12}$ M$_{\sun}$ as given by Finoguenov et
al.\ (2011), where $M_{200}$ is the total gravitational mass (assuming
$h=0.7$) within a radius where the average density is $200$ times the
critical density \cite[e.g.,][]{finoguenov01}. Several of the X-ray
groups are coincident with known overdensities in the CDFS
\citep[e.g.,][]{gilli03,adami05,trevese07,salimbeni09}, and we find
that the group members are preferentially found to have higher values
of $\log_{10}(1+\delta_3)$, thereby providing an independent check of
the environment measures presented here.

\begin{deluxetable*}{c c c c c c c c c c}
\tablewidth{0pt}
\tablecolumns{10}
\tablecaption{\label{tab_environ} ACES Environment Catalog}
\tablehead{\multirow{2}{*}{Object ID\tablenotemark{{\it a}}} &
  \multirow{2}{*}{$z$\tablenotemark{{\it b}}} & 
  \multirow{2}{*}{Ref\tablenotemark{{\it c}}} & 
  \multirow{2}{*}{$D_{\rm edge}$\tablenotemark{{\it d}}} & \multicolumn{3}{c}{ACES only tracer} &
  \multicolumn{3}{c}{ACES + $z_{\rm public}$ tracer} \\ 
    &     &    &    &  $\log(1+\delta_{3})$\tablenotemark{{\it e}} &
    $\log(1+\delta_{4})$\tablenotemark{{\it f}} & $\log(1+\delta_{5})$\tablenotemark{{\it g}} &
    $\log(1+\delta_{3})$\tablenotemark{{\it e}} &
    $\log(1+\delta_{4})$\tablenotemark{{\it f}} & $\log(1+\delta_{5})$\tablenotemark{{\it g}} } 
\startdata
121 & 0.3164 & 0 & 0.077 & -0.930 & -0.768 & -0.679 & -0.975 & -0.831 & -0.731 \\ 
122 & 0.3869 & 0 & 0.055 & -0.128 &  0.028 &  0.018 & -0.169 & -0.050 & -0.041 \\ 
253 & 0.7689 & 0 & 0.125 & -0.489 & -0.313 & -0.250 & -0.542 & -0.360 & -0.305 \\ 
263 & 0.6023 & 0 & 0.081 & -1.601 & -1.434 & -1.416 & -1.617 & -1.478 & -1.371 \\ 
\vspace*{-0.1in}
\enddata
\tablecomments{Table \ref{tab_environ} is presented in its entirety in
  the electronic edition of the Journal. A portion is shown here for
  guidance regarding its form and content. Readers are reminded that
  galaxies located near a survey edge (e.g., within $\sim 1$ $h^{-1}$
  comoving Mpc) should be excluded from analyses, since the associated
  environment measures are contaminated by edge effects (see \S
  \ref{sec_environ}).}

\tablenotetext{{\it a}}{Object identification number (SEQ) in $R$-band
  catalog of \citet{wolf04}.}

\tablenotetext{{\it b}}{Redshift (in the heliocentric frame).}

\tablenotetext{{\it c}}{Source of redshift: (0) this work;
  (1) DEEP2/DEIMOS ; (2) \citet{lefevre04}; (3) \citet{vanzella05};
  (4) \citet{vanzella06}; (5) \citet{mignoli05}; (6)
  \citet{ravikumar07}; (7) \citet{szokoly04}; (8) \citet{balestra10}}

\tablenotetext{{\it d}}{Distance from nearest survey edge ($h^{-1}$ comoving
  Mpc).}

\tablenotetext{{\it e}}{Overdensity as given by the third-nearest-neighbor
  surface density.}

\tablenotetext{{\it f}}{Overdensity as given by the fourth-nearest-neighbor
  surface density.}

\tablenotetext{{\it g}}{Overdensity as given by the fifth-nearest-neighbor
  surface density.}
\end{deluxetable*}

\section{Summary and Future Work}
\label{sec_future}

We present a spectroscopic survey of the Chandra Deep Field South
(CDFS), conducted using Magellan/IMACS and aptly named the Arizona
CDFS Environment Survey (ACES). The survey dataset includes $7277$
unique spectroscopic targets, yielding $5080$ secure redshifts, within
the extended CDFS region. We describe in detail the design and
implementation of the survey and present preliminary redshift and
environment catalogs.

While this work marks a significant increase in both the spatial
coverage and the sampling density of the spectroscopic observations in
the CDFS, there remains much analysis of the ACES data to be completed
in the future. In particular, work is presently underway to produce a
relative throughput correction for the spectra, using observations of
F stars that were targeted on many of the IMACS slitmasks. The F stars
were observed with the same instrumental set-up as the science targets
(i.e., same slitwidth, slitlength, etc.) and were included on multiple 
slitmasks, such that the spectra fell on each of the $8$ IMACS CCDs,
allowing chip-to-chip variations in the throughput to be estimated. 

As discussed in \S \ref{sec_design}, fainter targets ($R \gtrsim
22.5$) were observed on multiple ($\sim \! 2$--$4$) slitmasks, with
the goal of accumulating longer integration times. In the future,
these data spanning different slitmasks will be combined, which will
likely improve the survey's redshift success rate at fainter
magnitudes. In parallel to this work, efforts to improve the spectral
reduction procedures are currently underway, which should likewise
improve the redshift completeness at all magnitudes. Likewise, future
work will include extracting spectra and measuring redshifts for
serendipitous detections of objects, an effort that could be helped
greatly by the addition of data from different slitmasks. The
completion of this ongoing work is expected to coincide with a final
data release, including updated redshift and environment catalogs as
well as all of the reduced IMACS spectra. Finally, analysis is
underway to utilize the ACES data to study the correlations between
star-formation history, morphology, and environment at $z < 1$, using
the rich multiwavelength data in the CDFS and the increased sample
size to improve upon previous efforts \citep[e.g.,][]{elbaz07,
  capak07, cooper08a, cooper10a}.

\vspace*{0.25in} 

\acknowledgments This work is based in part on observations made with
the Spitzer Space Telescope, which is operated by the Jet Propulsion
Laboratory, California Institute of Technology under a contract with
NASA. Support for this work was provided by NASA through the Spitzer
Space Telescope Fellowship Program. MCC acknowledges support provided
by NASA through Hubble Fellowship grant \#HF-51269.01-A, awarded by
the Space Telescope Science Institute, which is operated by the
Association of Universities for Research in Astronomy, Inc., for NASA,
under contract NAS 5-26555. MCC also acknowledges support from the
Southern California Center for Galaxy Evolution, a multi-campus
research program funded by the University of California Office of
Research. This work was also supported in part by NSF grant
AST-0806732 and by NASA through an award issued by JPL/Caltech as part
of the FIDEL Spitzer Legacy science program. We thank the DEEP2 Galaxy
Redshift Survey team for providing access to their Keck/DEIMOS
observations of the CDFS. MCC thanks John Mulchaey, Alexis Finoguenov,
and Dave Wilman for helpful discussions throughout much of the project
and also thanks the entire Las Campanas Observatory staff for their
help in the acquisition of the ACES Magellan/IMACS data. Finally, MCC
thanks Mike Boylan-Kolchin for helpful discussions and the anonymous
referee for valuable comments and suggestions that improved this work.

{\it Facilities:} \facility{Magellan:Baade (IMACS)}

\end{document}